%% file: main.tex
\documentclass[conference]{IEEEtran}
\usepackage{amsmath,amssymb,amsthm}
\usepackage{booktabs}
\usepackage{adjustbox}
\usepackage{graphicx}
\usepackage{xcolor}
\usepackage{tikz}
\usetikzlibrary{positioning}
\usepackage{cite}
\usepackage[hidelinks]{hyperref}

\newif\iffinal \finalfalse
\iffinal
  \newcommand{\todo}[1]{}
\else
  \newcommand{\todo}[1]{\textcolor{blue}{[TODO: #1]}}
\fi
\newtheorem{proposition}{Proposition}
\input{generated/macros}

\input{generated/macros_v3}

\input{generated/macros_fu}

\input{generated/macros_f2}
\input{generated/macros_mimic}
\input{generated/macros_f3}

\input{generated/macros_gbm2}

\ifdefined\SyntheticData
  \usepackage[text={SYNTHETIC DATA -- NOT RESULTS},scale=0.45,color=red!35]{draftwatermark}
\fi
\newcommand{\CR}{C^{R}}

\begin{document}

\title{Who Gets Re-ranked When a Clinical Model Gains New Data?}

\author{\IEEEauthorblockN{Priyanka Siddappa, Muntaqim Ahmed Raju, Shifat Haider Al Amin, and Ruizhe Ma}
\IEEEauthorblockA{Miner School of Computer \& Information Sciences, University of Massachusetts Lowell\\
Lowell, MA 01854, USA\\
Priyanka\_Siddappa@student.uml.edu, MuntaqimAhmed\_Raju@student.uml.edu\\
MdShifat\_AlAmin@student.uml.edu, Ruizhe\_Ma@uml.edu}}

\maketitle

\begin{abstract}
Hospitals often update a deployed risk model by adding a data source, such as laboratory
results, that only some patients have. Work on safe model updates assumes fixed inputs
and so ignores patients who lack the new data. The update has no new measurements for
them, yet their rank can change, and a sick patient not yet tested can drop off an alert
list. We split rank-based compatibility by data availability and pair
it with error correction and alert-list effects. We compared each update with ordinary
retraining on the old inputs, predicting ICU admission or death and hospital admission, on \NVisits{} emergency department visits from MC-MED and,
in a pre-specified replication, \MimN{} stays from MIMIC-IV-ED at a second hospital.
Radiograph report text gave no detectable gain and almost no extra re-ranking.
Laboratory results improved prediction but, at both hospitals and with linear or tree
models, pushed patients without labs down alert lists; linear updates also reshuffled
them more than retraining did, without ranking them better. A top-20\% admission list caught \ETwentyMmSensU{} of
admitted patients without labs after the update, against \ETwentyReSensU{} after
retraining (second hospital: \MimAdMmSUTwenty{} against \MimAdReSUTwenty), while catching
more patients overall. This held for list sizes from 5\% to 50\% and in every patient group we checked, despite better measured calibration. Separate alert thresholds for patients with and without labs restored most or all of their catch
rate at both hospitals at almost no cost. Teams should check these patients before deploying such an update.
\end{abstract}

\begin{IEEEkeywords}
model updating, backward compatibility, multimodal learning, missing modality,
emergency medicine, informative missingness
\end{IEEEkeywords}

\section{Introduction}
Hospitals keep updating their prediction models: they retrain them as patients and
practice change, and more and more often they add new kinds of data. Better overall accuracy does not guarantee that an update keeps doing
what the old model did well, and new errors on cases the old model got right can hurt
users, such as clinicians, who have learned to work with it~\cite{bansal2019updates}. For risk
stratification, \"Otle\c{s} et al.~\cite{otles2023updating} defined \emph{rank-based
compatibility} $\CR$: the probability that the updated model orders a pair of patients
with different outcomes correctly, given that the original model did. Adding a
differentiable version of $\CR$ to the training loss made their updates more compatible
at little cost in AUROC.

That work, like most studies of update
stability~\cite{otles2023updating,srivastava2020empirical,trauble2021backward,
lopezmartinez2022instability,bilionis2026empirical}, keeps the model's inputs fixed.
Many real updates do not. They add a data source that exists only for patients for whom
someone ordered it. Ordering follows clinical suspicion and local
practice~\cite{agniel2018biases,groenwold2020informative}, and it can differ across
patient groups~\cite{chang2022disparate}.

\begin{figure}[t]\centering
\input{figs/fig_concept}
\caption{Test pairs grouped by who has the added data. For linear updates, UU-pair swaps reflect re-estimated weights on old features.}
\label{fig:concept}
\end{figure}
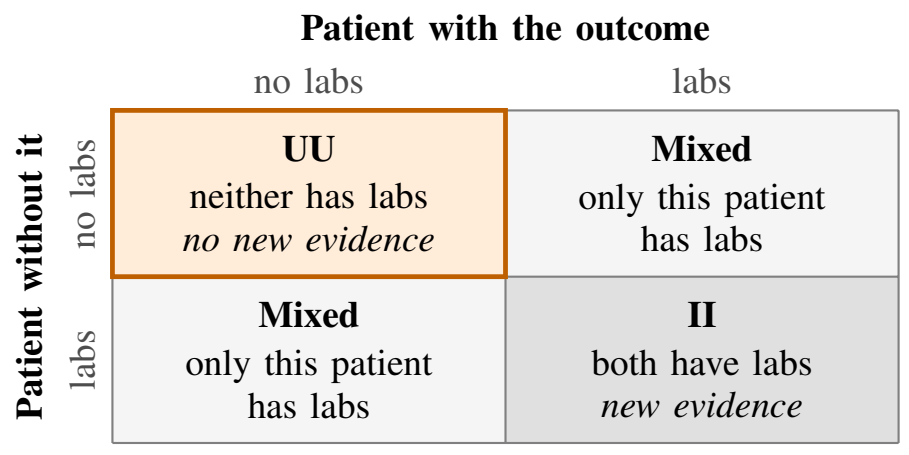

Adding data this way creates an asymmetry (Fig.~\ref{fig:concept}). If the updated
model swaps two patients who both have the new data, the swap may reflect new evidence.
If it swaps two patients who both lack it, it cannot reflect new measurements for either
of them. For a linear update, their score difference depends on re-estimated weights on
the original features. The availability indicator can still affect those weights during
training, even though it has the same value for both patients at evaluation. Aggregate
$\CR$ mixes these cases together.

This matters in practice. An emergency department (ED) can use risk scores to decide
whom to look at first, for example by reviewing the highest-risk visits on a list. A
patient who has had no laboratory tests gets no new measurements from an update that adds
laboratory results, but still competes for a place on that list with patients who now score
higher. If the update pushes such patients down, clinicians see their relative risk
change with nothing to explain it, and a seriously ill patient who has not yet been
tested can drop off the list.

We ask three questions. When an update adds data that only some patients have, are the
patients without it re-ranked more than ordinary retraining would re-rank them? Do they
lose their place on alert lists? And can that be prevented without giving up what the
update gains? We compare with retraining on the old inputs, because retraining alone
already reshuffles some patients.

In this paper we:
\begin{itemize}
\item split $\CR$ by who has the added data and report it with error correction and
alert-list effects, and show (Prop.~\ref{prop:uu}) how patients without the new data
can move at all;
\item show, in three simulated updates on \NVisits{} ED visits from
MC-MED~\cite{kansal2025mcmed}, that patients without the new data were re-ranked more
only when the new data actually helped prediction, in two pre-specified tests and a
time-ordered split, but not with gradient-boosted trees;
\item show that the update that helped prediction pushed these patients down alert
lists. We first saw this in an exploratory analysis and then confirmed it in a
pre-specified test. The loss persists despite better measured calibration, holds at list sizes from 5\% to 50\%, over
time, in every race, age, sex and payer group and with gradient-boosted trees, and disappears when the
added data is random noise;
\item compare eight ways to prevent this. Fixes that protect these patients' order help
only a little. Separate alert thresholds for patients with and without the data undo
most or all of the alert loss at almost no cost;
\item repeat the main tests on \MimN{} ED stays from a second hospital
(MIMIC-IV-ED~\cite{johnson2023mimicived}), with four hypotheses written down before any
model was trained there. All four held, with a near-identical alert loss and fix.
\end{itemize}

\section{Related Work}
Several papers ask what happens to the people who use a model when the model is
updated. Bansal et al.~\cite{bansal2019updates} introduced backward compatibility of AI
updates with users' mental models. Srivastava et al.~\cite{srivastava2020empirical}
studied empirically when updates break it, and Tr\"auble et al.~\cite{trauble2021backward}
proposed a probabilistic way to update stored predictions while limiting new errors. \"Otle\c{s} et al.~\cite{otles2023updating}
introduced $\CR$ and a compatibility loss for clinical risk models.
Positive-congruent training~\cite{yan2021pct} reduces new mistakes by teaching the new
model to copy the old one where the old one was right. As far as we know, none of these
papers looks at an update that adds a new input.

Other work studies how unstable clinical models are when they are retrained.
Lopez-Martinez et al.~\cite{lopezmartinez2022instability} found that retraining a deep
clinical model on the same data changed patient-level predictions but not AUROC. Bilionis et al.~\cite{bilionis2026empirical} found that repeated
retraining on the same features changes predictions more for some sociodemographic
groups, which overall metrics can miss. Shen et al.~\cite{shen2024data} showed that adding training \emph{samples} from
new sources can lower accuracy and worst-group performance. We look at the
other kind of addition: a new \emph{input}.

Many methods handle data with missing parts. Some learn with missing
modalities~\cite{ma2021smil,reza2025robust}, some combine EHR data with
imaging~\cite{hayat2022medfuse,jabbour2022combining}, and others, such as
M3Care~\cite{zhang2022m3care} and modality dropout~\cite{neverova2016moddrop}, are built
to stay accurate when a modality is absent. None of them compares the new model with
the one it replaces. A recent EHR and chest radiograph
benchmark~\cite{yin2026carebench} found that fusion gains fade when radiographs are
missing, unless the model is built for incomplete inputs, and that fusion does not by itself close gaps between groups. In EHR data, whether a test was done at
all carries information~\cite{agniel2018biases,groenwold2020informative}, and when some
groups are tested less, models trained on test-based labels can end up biased against
them~\cite{chang2022disparate,chang2024biased}.

Our alert analysis is close to work on fairness in rankings. xAUC~\cite{kallus2019xauc}
measures how often the positives of one group are ranked above the negatives of
another, and fairness of exposure~\cite{singh2018exposure} asks how attention is shared
in a ranking. We ask a similar question about patients with and without a data source,
across an update. The fix we test, separate thresholds per group, follows the post-processing idea of
Hardt et al.~\cite{hardt2016equality}, who set group-specific thresholds to equalize
true positive rates. Here the groups are defined by data availability, and the target is each
group's share of alerts. Conformal risk control~\cite{angelopoulos2024crc} could, under its assumptions,
set such thresholds with a guarantee on a chosen loss; we leave this to future work.

\section{Problem Setup and Method}\label{sec:method}
\subsection{Rank-based compatibility and error correction}
Each patient $i$ has original features $\mathbf{x}_i$, a binary outcome $y_i$, an
indicator $h_i\in\{0,1\}$ of whether the added data is available, and, if $h_i=1$, added
features $\mathbf{z}_i$. The original model $f^o(\mathbf{x})$ uses only $\mathbf{x}$; the
updated model $f^u(\mathbf{x},h,\mathbf{z})$ also uses the added data. Let
$\mathcal{P}=\{(i,j):y_i=0,y_j=1\}$, and let $o_{ij}=\mathbf{1}[f^o_i<f^o_j]$ and
$u_{ij}=\mathbf{1}[f^u_i<f^u_j]$ indicate correct ordering by each model. For a set of
pairs $\mathcal{T}\subseteq\mathcal{P}$,
\begin{equation}
\CR_{\mathcal{T}}=\frac{\sum_{\mathcal{T}} o_{ij}u_{ij}}{\sum_{\mathcal{T}} o_{ij}},\qquad
\mathrm{IMP}_{\mathcal{T}}=\frac{\sum_{\mathcal{T}} (1-o_{ij})u_{ij}}{\sum_{\mathcal{T}} (1-o_{ij})}.
\end{equation}
With $\mathcal{T}=\mathcal{P}$, $\CR$ is the measure of~\cite{otles2023updating}. IMP is
the fraction of the original model's misordered pairs that the update orders correctly.
An update that copies the original model reaches $\CR=1$ while correcting nothing, and
IMP makes that visible. We call $1-\CR_{\mathcal{T}}$ the \emph{flip rate}.

\subsection{Decomposition by data availability}
We partition $\mathcal{P}$ into $\mathcal{P}_{UU}$ (neither patient has the added data),
$\mathcal{P}_{II}$ (both have it) and $\mathcal{P}_{MIX}$ (exactly one has it), and we
write $\mathcal{P}_{UU,g}$ for pairs in $\mathcal{P}_{UU}$ whose patients both belong to
demographic group $g$. A re-ordering in $\mathcal{P}_{II}$ or $\mathcal{P}_{MIX}$ may
reflect new evidence. One in $\mathcal{P}_{UU}$ cannot use new values.

The next observation is simple, but it names the one route by which such
re-orderings can arise.
\begin{proposition}\label{prop:uu}
Let $f^u(\mathbf{x},h,\mathbf{z})=\mathbf{w}_x^\top\mathbf{x}+w_h h+\mathbf{w}_z^\top(h\,\mathbf{z})+b$.
For any $(i,j)\in\mathcal{P}_{UU}$, $u_{ij}=\mathbf{1}[\mathbf{w}_x^\top(\mathbf{x}_j-\mathbf{x}_i)>0]$.
Hence, if the original model is also linear, $f^o=\mathbf{v}^\top\mathbf{x}+c$, every flip
in $\mathcal{P}_{UU}$ results from the difference between $\mathbf{w}_x$ and $\mathbf{v}$.
\end{proposition}
\begin{proof}
With $h_i=h_j=0$, the $w_h$ and $\mathbf{w}_z$ terms vanish for both patients, and $b$
cancels in $f^u_j-f^u_i$.
\end{proof}
When the updated model is fit, $\mathbf{w}_x$ is estimated jointly with $\mathbf{w}_z$.
Original features whose signal overlaps with the added data can lose weight, and these
re-estimated weights are then applied to patients who do not have the added data. We
therefore hypothesize, and test empirically, that flips in $\mathcal{P}_{UU}$ exceed those of
ordinary retraining only when the added data is informative.

\subsection{Ways to reduce re-ranking}
We compare three methods here and add more in Sec.~\ref{sec:extra}. The first, RBC,
is the method of~\cite{otles2023updating}. It minimizes
$\alpha\mathcal{L}_{\mathrm{BCE}}+(1-\alpha)(1-\tilde{C}^R_{\mathcal{P}})$, using the
smooth surrogate $\tilde{C}^R_{\mathcal{T}}=\sum_{\mathcal{T}}\sigma(s\Delta^o_{ij})\sigma(s\Delta^u_{ij})/
\sum_{\mathcal{T}}\sigma(s\Delta^o_{ij})$ with $\Delta_{ij}=f_j-f_i$. We compute $\Delta$ on
the logit scale, whereas~\cite{otles2023updating} used predicted probabilities. Both
target the same $\CR$, which depends only on order, but they weight pairs differently
during training (on the probability scale, pairs of two low-risk patients have small
score differences), so the solutions found can differ. For that reason we also
report the probability-scale version (Sec.~\ref{sec:results}).
The second, MARBC, is our own variant of RBC. It puts weight $\lambda$ on pairs without
the added data, averaged over racial groups so that small groups count equally.
The third, routing, keeps the original model's score for patients without the added data and
uses the updated model's score otherwise, so $\CR_{UU}=1$ and $\mathrm{IMP}_{UU}=0$ by
construction.

\section{Experimental Setup}
\subsection{Data and features}
We use MC-MED~\cite{kansal2025mcmed,kansal2025mcmeddata}, a credentialed-access dataset of
\NAllVisits{} adult ED visits to one academic medical center between 2020 and 2022, with triage
data, minute-level vital signs, laboratory results and radiology results. We make
predictions $T=3$\,h after arrival, when many patients are still being worked up and
some have no laboratory results yet. We keep visits whose ED disposition was decided
after $T$, so that the prediction is made before the outcome is known, which leaves \NVisits{} visits from \NPatients{} patients. Groups are
self-reported race (Asian, Black, Native Hawaiian or Pacific Islander [NHPI], White,
Other/Unknown). We use them for evaluation and inside MARBC, never as model inputs.

The triage features are age, sex, the number of earlier visits in the
dataset, acuity (Emergency Severity Index), six triage vital signs and mode of arrival.
Ten monitored measures, such as heart rate, blood pressure and oxygen saturation, are
summarized up to $T$ by their mean, minimum, maximum, last value and count. The
laboratory features are the 25 components most often resulted by $T$, with the same
summaries. We median-impute missing values, add missingness indicators and standardize,
fitting these on each model's training data, except that the added lab block was imputed and scaled
on all visits (no labels); Tab.~\ref{tab:sens} gives a training-only check.

\subsection{Three experiments}
Each experiment compares two updates trained on the same data. The \emph{EHR-only
retrain} keeps the original model's inputs, so it shows how much retraining alone
reshuffles patients. The \emph{multimodal update} adds the new source and an indicator
of whether a patient has it. \emph{Experiment~1} predicts ICU admission from the ED or death
before discharge, which occurred in \OutcomeRate{} of visits (\NPos{} events). The
original model uses triage, monitoring and laboratory features. The update adds an
availability indicator and the radiologist's impression of any chest radiograph
resulted by $T$, encoded simply with TF-IDF and SVD (see Limitations) and reduced to
32 dimensions by PCA fit on development patients. \emph{Experiment~2} keeps the outcome. Its original model
uses triage and monitoring features, and the update adds an availability indicator and
the laboratory features for patients with at least one of the 25 components by $T$,
reduced to 64 dimensions by PCA fit on development patients. Before running it, we wrote
down the hypothesis that flips among pairs without labs would be at least 0.5
percentage points (pp) higher for the multimodal update than for an EHR-only retrain,
with a 95\% CI excluding 0. \emph{Experiment~3} repeats Experiment~2 with hospital
admission as the outcome (\EAdmitRate{} of visits, \EAdmitN{} events), which gives many
more events among patients without labs. After the exploratory alert analysis of
Experiment~2, and before running Experiment~3, we recorded two hypotheses: the same
0.5\,pp re-ranking criterion, and a lower catch rate of admitted patients without labs
on a top-20\% alert list for the multimodal update than for the EHR-only retrain.

\subsection{Training and evaluation}
We follow the setup of~\cite{otles2023updating} at a smaller scale. Each repetition
randomly assigns patients, keeping all of a patient's visits together, to an
original-model set (15\%), an update set (55\%, split equally into development and
validation) and a test set (30\%). The original model is L2-regularized logistic
regression, trained on half of its set with the regularization strength chosen on the
other half. As in~\cite{otles2023updating}, it is trained on less data than the updates,
like an older model built before more data was collected. Updates are linear and trained with L-BFGS for up to 400 iterations. For the
EHR-only retrain and for the multimodal update trained with cross-entropy (BCE), we
build pools of \NCand{} candidates: three regularization strengths on the full
development set and six bootstrap resamples. RBC and MARBC start from the full-data
multimodal BCE model with the highest validation AUROC, use its regularization
strength, and try $\alpha\in\{0,0.25,0.5,0.75\}$, with $\lambda=\LamUU$, $s=2$ and
40,000 sampled pairs. For each method we keep the model that maximizes
$\beta\,\mathrm{AUROC}+(1-\beta)\,\CR$ on validation data. We report $\beta=\Beta$ and
use $\beta\in\{0.3,0.7\}$, and $\CR$ computed on UU pairs, as sensitivity analyses.
On the test set we report AUROC, AUROC among patients without the added data, $\CR$ and
IMP for each pair type, and the alert-list results described below. Across \NReps{}
repetitions we report paired differences with 95\% $t$-intervals and Wilcoxon
signed-rank tests. These repeated-split intervals describe variability conditional on
this cohort, not uncertainty across independently sampled clinical populations. All
models are fit with NumPy and SciPy, except the boosted trees
(scikit-learn). Only the pre-specified tests are confirmatory. For the three
of them (one in Experiment~2, two in Experiment~3), the largest Holm-adjusted $p$-value
is \HolmMax{}. Every other comparison is descriptive and not adjusted for multiplicity.

\subsection{Two extra checks}
The first check uses time. MC-MED provides a
chronological split in which later parts hold later visits and no patient appears in
two parts~\cite{kansal2025mcmeddata}. We train the original model on the oldest visits,
retrain the update on everything available later (the oldest visits plus a random half
of the validation-period patients), select on the other half, and test on the newest
visits. We run \TNReps{} repetitions on this fixed test set, so the intervals reflect only
training variation. The second check uses a different kind of model: both updates are scikit-learn's
LightGBM-style~\cite{ke2017lightgbm} gradient-boosted trees, trained with early stopping over the same
\NReps{} random splits. The multimodal one receives raw laboratory values and handles
missing values natively. We also ran this check, with alert lists and two thresholds,
on Experiment~3 and both outcomes at the second hospital, which have far more events.

\subsection{Alert lists}
Each model flags its own highest-risk $k$\% of test visits, so every model and policy
produces the same number of alerts and the same workload. Among
visits without the added data, we report the share flagged and the share of
outcome-positive visits flagged, which we call \emph{caught}. This analysis was
exploratory in Experiment~2 ($k\in\{5,10\}$) and pre-specified in Experiment~3 ($k=20$
primary, $k=10$ secondary).

\subsection{Additional analyses}\label{sec:extra}
To find out why patients without labs lose alerts and how to prevent it, we ran further
analyses on Experiments~2 and~3 with the same splits, candidate pools and selection rule.
We rebuilt the selected EHR-only retrain and multimodal update for every repetition, and
their test AUROC matched the main analysis exactly. Five variants of the update help
separate the causes:
\begin{itemize}
\item lab values without the availability indicator, so that absent labs sit at the
lab-block mean;
\item the indicator without lab values;
\item random Gaussian features in place of lab values for patients with labs, without
the indicator, as uninformative added data with the same availability;
\item a version of modality dropout~\cite{neverova2016moddrop} that hides labs and indicator for a
random 30\% of development patients with labs;
\item a \emph{frozen} original model plus a trained additive term on the indicator and
lab values, which leaves the order of patients without labs unchanged.
\end{itemize}
We also ran RBC with the surrogate computed on predicted probabilities, as
in~\cite{otles2023updating}, with $s\in\{10,50\}$, and an update whose weights on the
original features are pulled toward the retrain's by a penalty
$\gamma\lVert\mathbf{w}_x-\mathbf{w}_x^{\mathrm{retrain}}\rVert^2$
($\gamma\in\{0.1,1,10\}$). Last, we tested a fix that acts on the alert list instead of the
model: a \emph{two-threshold} alert policy. It keeps the update's scores but fixes the share of alerts that go to visits without labs
at the retrain's share on validation data, and fills the rest of the list with visits
with labs. We also varied this share from zero to twice the retrain's share.

To check that the alert loss is not an artifact of how we measured it, we report
calibration among visits without labs, both as observed over expected events (O/E) and
as expected calibration error (ECE; 10 equal-frequency bins), overall and by race, catch rates for
list sizes from 5\% to 50\%, and catch rates by self-reported race, age band, sex and
payer. We also replaced the top-$k$\% list with a fixed risk cutoff (where the retrain
flags $k$\% of validation visits); there the two-threshold idea becomes two cutoffs,
with visits without labs getting the cutoff at which the update matches the retrain's
validation flag rate among them. We repeated Experiment~3 on the time-ordered split
(\PTNReps{} repetitions; both candidate pools trained and selected with the main rule).
Finally, we report xAUC between availability groups~\cite{kallus2019xauc}, here the
probability that an outcome-positive visit without labs scores above an outcome-negative
visit with labs, and the weights each model puts on the original features.

\subsection{Second hospital}\label{sec:mimicsetup}
To see whether the lab results hold elsewhere, we repeated Experiments~2 and~3 on
MIMIC-IV-ED 2.2~\cite{johnson2023mimicived}, linked to MIMIC-IV
3.1~\cite{johnson2023mimiciv,johnson2024mimiciv31}. These come from the ED of a different
academic medical center, in another US city. We kept adult stays still in the ED at
$T=3$\,h (\MimAll{} stays from \MimPatients{} patients) and, to keep compute close to
MC-MED, drew \MimN{} of them at random by patient. The features follow MC-MED where the
data allow: age, sex, earlier ED stays, acuity, six triage vital signs, arrival mode, and
the same summaries of six charted vital signs up to $T$. A lab result counts as
available if the sample was taken after arrival and the result was stored by $T$. We use
the 25 most common components. The outcomes are ICU admission within 12\,h of leaving
the ED or death in hospital (\MimIcuRate{} of stays), and admission from the ED
(\MimAdRate). Code and settings are those of the main analysis, including the
two-threshold policy.

We wrote down four hypotheses after seeing only cohort counts and outcome rates, before
building lab features or training any model on these data. R1 and R2: for each outcome,
flips among pairs without labs are at least 0.5\,pp higher for the update than for the
retrain, with a 95\% CI excluding 0. R3: on a top-20\% admission list, the update catches
fewer admitted stays without labs than the retrain. R4: with two thresholds, the catch
rate for admitted stays without labs is within 3\,pp of the retrain's, and the catch rate
for all admitted stays is higher. We correct the four with Holm's method, counting R4 as
one test that needs both parts.

\section{Results}\label{sec:results}

In the tables, Flipped is the \% of correctly ordered UU pairs that the new model misorders and
Fixed the \% of misordered ones it corrects; Caught is the \% of outcome-positive visits on the
alert list and Prec.\ the \% of alerts that are outcome-positive; U and I mark visits without and
with labs, Share U is the \% of alerts given to U visits, and AUROC$_\varnothing$ is AUROC among
patients without the added data. Results are means over 10 repetitions, except the time-ordered analyses, which use 6.

\input{generated/tab_cohort}

\subsection{Who has the added data}
Sicker patients were more likely to have the added data, and availability differed by
race. Chest radiograph results were available by 3\,h for \XRate{} of visits: \XRatePos{} of patients with the
ICU/death outcome and \XRateNeg{} of those without it, ranging from \XMin{}
(\XMinGroup) to \XMax{} (\XMaxGroup) across groups ($\chi^2_{\XChiDof}=\XChi$, \XChiP).
Laboratory results were available for \LRate{} of visits: \LRatePos{} of patients with the
ICU/death outcome, \ELabPos{} of admitted patients and \ELabNeg{} of patients not
admitted, ranging from \LMin{} (\LMinGroup) to \LMax{} (\LMaxGroup;
$\chi^2_{\LChiDof}=\LChi$, \LChiP; Tab.~\ref{tab:cohort}).

\input{generated/tab_summary}

\begin{figure*}[t]\centering
\includegraphics[width=\textwidth]{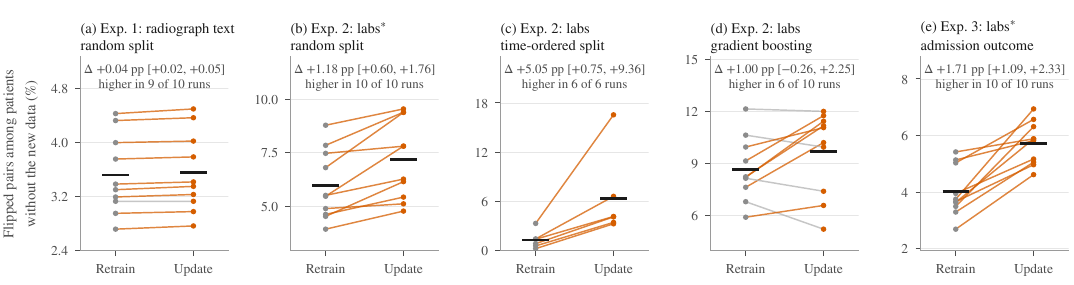}
\caption{Flipped UU pairs, retrain vs.\ update; one line per repetition. $^\ast$Pre-specified.}
\label{fig:forest}
\end{figure*}

\subsection{Q1: Are patients without the new data re-ranked more?}
With linear updates, patients without the new data were re-ranked more only when the new data helped
prediction. Tab.~\ref{tab:summary} and Fig.~\ref{fig:forest} show every setting. Radiograph
report text, as we encoded it, left AUROC where it was (\VXAUROCe{} vs.\ \VXAUROCm), and
patients without a radiograph were re-ranked almost exactly as often as under ordinary
retraining (\VXFe{} vs.\ \VXFm{} of pairs flipped; difference \VXX\,pp [\VXXLo,
\VXXHi]). Laboratory results raised AUROC in every setting, and they raised flips among
patients without labs: by \VLX\,pp [\VLXLo, \VLXHi] in Experiment~2 (from \VLFe{} to
\VLFm, a \VLRel{} increase) and by \VAX\,pp [\VAXLo, \VAXHi] in Experiment~3 (from
\VAFe{} to \VAFm). Both pre-specified criteria were met. The interval excluded 0 under
all \EHOneNSettings{} selection settings in Experiment~3 and under \LHOneNSig{} of
\LHOneNSettings{} in Experiment~2. The time-ordered split showed a larger effect
(\VTX\,pp [\VTXLo, \VTXHi]), though its test period holds only \TNPosU{} critically ill
visits without labs, so this estimate is fragile. Gradient boosting showed little of this. Its Experiment~2 interval was wide and included 0
(\VGX\,pp [\VGXLo, \VGXHi]), and boosted updates did not reshuffle more than a boosted
retrain in Experiment~3 (\GcFlip\,pp [\GcFlipLo, \GcFlipHi]) or for admission at the second hospital (\GmFlip\,pp
[\GmFlipLo, \GmFlipHi]); for ICU/death there the excess was small (\GiFlip\,pp [\GiFlipLo, \GiFlipHi]), below our 0.5\,pp bar.
The extra reshuffling is mainly a property of the linear updates that Prop.~\ref{prop:uu} describes;
trees can split patients by lab availability. In no setting did the
update order patients without labs any better. It ordered them worse in Experiment~3
($\Delta$AUROC$_\varnothing$ \VADAu{} [\VADAuLo, \VADAuHi]) and in the time-ordered split
(\VTDAu{} [\VTDAuLo, \VTDAuHi], on few events).

\input{generated/tab_decomp}

Some of this re-ranking fixed old errors (Tab.~\ref{tab:decomp}). In Experiment~3 the update corrected \DAFixUUm{} of the original model's misordered UU
pairs, compared with \DAFixUUe{} for retraining. Among pairs in which both patients had
labs it corrected \DAFixIIm{}, compared with \DAFixIIe. So the reshuffling among patients without labs
is not pure noise, but for them the net result is no better, or slightly worse. The
weights move as Prop.~\ref{prop:uu} suggests. In Experiment~3, \CoefShrink{} of the
weights on the original features shrank after the update, and the largest change was the weight on age, which fell from \CoefAgeRe{} to
\CoefAgeMm{} on the standardized scale. Older patients were more likely to have labs
(\LabAgeOld{} of visits at 65 or older, \LabAgeYoung{} under 45), so part of what age
used to signal probably now comes from the labs, and patients without labs lose that
part.

\subsection{Q2: Do they lose their place on alert lists?}
When the labs helped prediction, the update pushed patients without labs down the alert list. In the exploratory analysis of Experiment~2, a top-5\% list
caught \AFiveMMBCESensU{} of critically ill visits without labs after the update and
\AFiveEHRBCESensU{} after retraining (\AFiveDSensU\,pp [\AFiveDSensULo,
\AFiveDSensUHi]), while the catch rate for those with labs changed by \AFiveDSensI\,pp. The pre-specified test in
Experiment~3 confirmed this with much tighter intervals. A top-20\% list caught
\ETwentyMmSensU{} of admitted visits without labs after the update and \ETwentyReSensU{}
after retraining (\ETwentyDSensU\,pp [\ETwentyDSensULo, \ETwentyDSensUHi],
\ETwentyDSensUP), and it flagged \ETwentyMmRateU{} of all visits without labs instead of
\ETwentyReRateU, while the catch rate for admitted visits with labs changed by
\ETwentyDSensI\,pp. Across repetitions, each test set contained about \CntMcN{} visits,
including about \CntMcPos{} admitted visits without labs. The update flagged about
\CntMcMm{} of these visits, compared with about \CntMcRe{} after retraining. Fixing
the list size is not what causes this. At a fixed risk cutoff the update flagged more
visits than the retrain (\FcCutMmPct{} against \FcCutRePct), yet caught only
\FcCutMmSU{} of admitted patients without labs, against \FcCutReSU. Boosted trees lost these
patients too: at top 20\% the catch rate for admitted visits without labs fell from
\GcReSU{} to \GcMmSU{} (\GcDSU\,pp [\GcDSULo, \GcDSUHi]), and at the second hospital from
\GmReSU{} to \GmMmSU. Two thresholds restored their catch rate (\GcTwoSU{} and \GmTwoSU). The update caught more admitted patients
overall, not fewer: for every admitted patient without labs it stopped flagging, it
flagged about \ETradeRatio{} more admitted patients with labs. What changed was who got
the alerts. The time-ordered split of Experiment~3, with about 300 admitted visits
without labs in its test period, gave the same picture: flips among patients without labs differed by
\PTMmVsReFlip\,pp [\PTMmVsReFlipLo, \PTMmVsReFlipHi], and the top-20\% list caught
\PTMmSU{} of admitted visits without labs instead of \PTReSU{} (\PTMmVsReU\,pp
[\PTMmVsReULo, \PTMmVsReUHi]). xAUC tells the same story without fixing a list size. After retraining,
an admitted patient without labs scored above a non-admitted patient with labs
with probability \PAReXA; after the update, \PAMmXA, not far from a coin flip.

\begin{figure*}[t]\centering
\includegraphics[width=\textwidth]{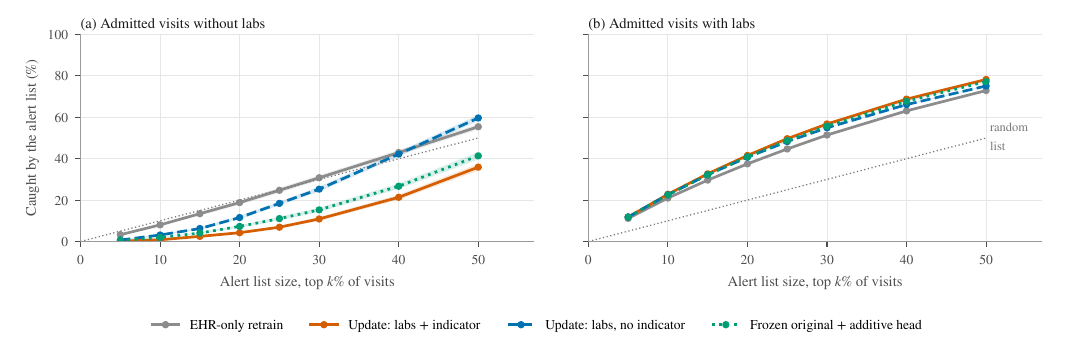}
\caption{Experiment~3: admitted visits caught by top-$k$\% lists (mean, 95\% CI). Dotted: random list.}
\label{fig:ksweep}
\end{figure*}

\input{generated/tab_ablation}

\subsection{Why they lose alerts}
In Experiment~3 the loss showed up at every list size from 5\% to 50\% (Fig.~\ref{fig:ksweep}), and most
of it came from the indicator of whether labs were available (Tab.~\ref{tab:ablation}).
After retraining, admitted patients without labs were caught at about the rate a random
list of the same size would achieve. After the update they were caught much less often,
whether the list was short (top 10\%: \KReUTen{} vs.\ \KMmUTen) or long (top 50\%:
\KReUFifty{} vs.\ \KMmUFifty).

The ablations pull apart two effects. Random features with the same availability and no
indicator behaved just like retraining (AUROC \EThrNoiseAuc, \EThrNoiseFlip{} of UU
pairs flipped, \EThrNoiseSens{} caught at top 20\%), as Sec.~\ref{sec:method} predicts
for uninformative data. The indicator on its own barely improved discrimination (AUROC
\EThrFlagOnlyAuc). Empirically, the indicator-only ablation had a UU flip rate of
\EThrFlagOnlyFlip{}, at the retrain level. This result is not guaranteed by the indicator
being constant within UU pairs: it can still change other coefficients during training.
It still cut the catch
rate to \EThrFlagOnlySens{} (\EThrFlagOnlyDSens\,pp [\EThrFlagOnlyDSensLo,
\EThrFlagOnlyDSensHi] relative to retraining). Lab values without the indicator did the
reverse: they produced the full rise in UU flips (\EThrNoFlagFlip, as for the full
update) and a smaller alert loss (\EThrNoFlagSens{} caught; \EThrNoFlagDSens\,pp
[\EThrNoFlagDSensLo, \EThrNoFlagDSensHi]). In this experiment, excess UU re-ranking appeared when lab values were included, not with
the indicator alone, consistent with re-estimated weights on original features. The alert loss in this
experiment comes mainly from the model learning that a patient who has not been tested
is probably at lower risk.

Experiment~2 did not follow that last pattern. There the indicator alone hardly moved
the catch rate (\ETwoFlagOnlySens{} vs.\ \ETwoReSens{} after retraining), while lab
values without it reproduced the full update (\ETwoNoFlagSens{} vs.\ \ETwoMmSens). How much of the alert loss
runs through the indicator depends on the outcome and the data, and Experiment~2, with
about \ANPosU{} relevant events per test set, is the noisier of the two.

The update did not underestimate these patients' risk. Of the visits without labs, \EThrMmObs{} ended in admission. The EHR-only retrain predicted \EThrRePred{} on
average (O/E \EThrReOE) and the update \EThrMmPred{} (O/E \EThrMmOE), so the update moved
their scores toward the observed rate without overshooting it. Its ECE among these
visits was lower too (\FcEceMm{} against \FcEceRe), and lower in all \FcEceGroups{}
racial groups. The loss arose despite better measured calibration. The update ranks patients partly by whether they have been tested, and it
orders the untested group no better than before (AUROC among visits without labs
\EThrReAucU{} vs.\ \EThrMmAucU).

\input{generated/tab_groups}

\subsection{Who loses alerts}
The alert loss was broad. At top 20\% in Experiment~3, the catch rate among admitted
visits without labs fell in all \GrpN{} racial groups, by \GrpRelMin{} to \GrpRelMax{} of
its value after retraining, while it rose for admitted visits with labs in every group
(Tab.~\ref{tab:groups}). The \GrpLowRe{} group started lowest, with only \GrpLowReV{}
caught even after retraining. The same held in every age band, both sexes and every
payer group (\SgN{} groups), with losses of \SgRelMin{} to \SgRelMax. The largest relative loss was for patients with
\SgWorst{} coverage, who were also caught least often to begin with (\SgWorstRe{} after
retraining, \SgWorstMm{} after the update).

\input{generated/tab_mitig}

\subsection{Q3: Can this be prevented?}
Mostly yes, but not by changing the model. Fixes that act on the model helped, but none
closed the gap (Tab.~\ref{tab:mitig}); a fix that acts on the alert list did
(Sec.~\ref{sec:twothr}). RBC, the compatibility loss of~\cite{otles2023updating}, reduced
flips among patients without the added data in all three experiments. Its AUROC cost was
not significant in Experiment~1 and small but significant in Experiments~2 and~3. It
also fixed fewer of the original model's errors, a cost that $\CR$ on its own hides.
Routing, by design, neither flipped nor fixed anything among patients without the added
data. Neither RBC nor routing restored the top-5\% catch rate in Experiment~2. Both
helped in Experiment~3 without reaching the retrain level of \ETwentyReSensU: RBC caught
\ETwentyRbcSensU{} and routing \ETwentyRtSensU{} of admitted visits without labs. Our variant MARBC
flipped more such pairs than RBC in every repetition, so we omit it from Tab.~\ref{tab:mitig}.

Two more baselines ran into the same wall (Tab.~\ref{tab:ablation}). Modality dropout
lowered AUROC to \EThrDropAuc{} in Experiment~3 and won back part of the loss
(\EThrDropSens{} caught). Freezing the original model and adding a lab term removed every UU
flip, as it must (AUROC \EThrAddAuc), yet caught only \EThrAddSens{} of admitted patients
without labs: their order stayed intact, but patients with labs moved past them.
Anchoring the update's weights to the retrain's did not help either. In our experiments,
it yielded a UU flip rate of \PAAncFlip{} (vs.\ \PAReFlip{} after retraining) and AUROC
of \PAAncAuc, but caught only \PAAncSU{} of admitted patients without labs
(Tab.~\ref{tab:policy}). RBC on the
probability scale, as in~\cite{otles2023updating}, changed little (Tab.~\ref{tab:ablation}).

\input{generated/tab_policy}

\subsection{A simple fix: separate thresholds}\label{sec:twothr}
Since the loss comes from patients without labs falling behind patients with labs, we
tried the most direct fix: keep the update's scores, but give visits without labs the
same share of the alert list they had after retraining (Tab.~\ref{tab:policy}). In
Experiment~3 this brought their catch rate back to \PATwoSU{}, against \PAReSU{} after
retraining (difference \PATwoVsReU\,pp [\PATwoVsReULo, \PATwoVsReUHi]). It kept almost
all of the update's overall gain: the list caught \PATwoSAll{} of all admitted patients,
against \PAMmSAll{} for the plain update and \PAReSAll{} for retraining. Precision barely
moved (\PATwoPPV{} vs.\ \PAMmPPV{} and \PARePPV). Raising the share from zero to twice the retrain's took the catch rate
without labs from 0 to \FcSwTwoU, while overall catch moved only from \FcSwZeroAll{} to
\FcSwTwoAll{} and precision stayed at \FcSwMinPPV{} or above. With
a fixed risk cutoff, two cutoffs did the same job (\FcCutTwoSU{} caught without labs). On the
time-ordered split the result was similar (\PTTwoSU{} vs.\ \PTReSU{} caught without
labs; \PTTwoSAll{} vs.\ \PTMmSAll{} overall). In Experiment~2 it recovered most of the
loss (\PLTwoSU{} vs.\ \PLReSU{} after retraining and \PLMmSU{} for the plain update),
again at almost no overall cost (\PLTwoSAll{} vs.\ \PLMmSAll). We also checked the policy within
each of the \TwoGrpN{} patient groups (race, age band, sex, payer) for admitted patients
without labs in Experiment~3. In \TwoGrpNBelow{} group the catch rate stayed clearly
below its retrain level: \TwoGrpWorst{} patients were caught \TwoGrpWorstTwo{} of the
time, against \TwoGrpWorstRe{} after retraining and \TwoGrpWorstMm{} with the plain
update (\TwoGrpWorstD\,pp [\TwoGrpWorstLo, \TwoGrpWorstHi] vs.\ retraining). In the other
groups the difference from retraining ranged from \TwoGrpOthMin{} to \TwoGrpOthMax\,pp.
Fixing the share of alerts for patients without labs does not by itself fix how that
share is split among them.

\input{generated/tab_mimic}

\begin{figure}[t]\centering
\includegraphics[width=\columnwidth]{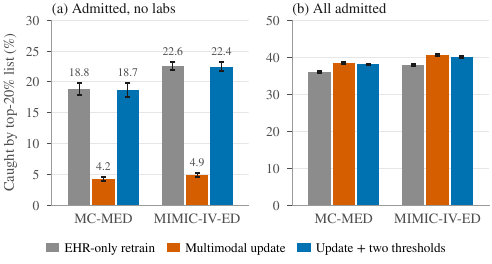}
\caption{Top-20\% admission list: \% caught (a) without labs, (b) overall.}
\label{fig:hospitals}
\end{figure}

\subsection{A second hospital}\label{sec:mimic}
All four pre-specified hypotheses held at the second hospital (largest Holm-adjusted
$p=\MimHolmMax$; Tab.~\ref{tab:mimic}, Fig.~\ref{fig:hospitals}).
Labs were available for \MimLab{} of stays, for \MimIcuLabPos{} of patients with the
ICU/death outcome, and for \MimAdLabPos{} of admitted patients against \MimAdLabNeg{} of
those not admitted. Across racial groups the share ranged from \MimLabMin{}
(\MimLabMinGroup) to \MimLabMax{} (\MimLabMaxGroup).

Labs raised AUROC for both outcomes and again ordered patients without labs slightly
worse ($\Delta$AUROC$_\varnothing$ \MimAdDAucU{} [\MimAdDAucULo, \MimAdDAucUHi] for
admission). The update flipped more pairs among these patients than retraining did:
\MimAdFlipD\,pp [\MimAdFlipDLo, \MimAdFlipDHi] for admission (R2) and \MimIcuFlipD\,pp
[\MimIcuFlipDLo, \MimIcuFlipDHi] for ICU/death (R1). The ICU/death estimate clears the
0.5\,pp bar we set, but the low end of its interval does not, so we read it as a small
effect. As before, the update fixed far more of the old model's errors than
retraining did among pairs with labs (\MimAdFixIIMm{} against \MimAdFixIIRe{} for
admission), but hardly more among pairs without (\MimAdFixUUMm{} against
\MimAdFixUURe), and \MimCoefShrink{} weights on the original features shrank. Age again changed most (\MimCoefAgeRe{} to \MimCoefAgeMm).

The alert loss came back at nearly the same size. A top-20\% admission list caught
\MimAdMmSUTwenty{} of admitted stays without labs after the update and \MimAdReSUTwenty{}
after retraining (\MimAdMmVsReU\,pp [\MimAdMmVsReULo, \MimAdMmVsReUHi]; R3), against
\ETwentyMmSensU{} and \ETwentyReSensU{} at MC-MED. In a test set of about \CntMiN{}
stays, that is \CntMiMm{} of \CntMiPos{} admitted stays without labs instead of \CntMiRe. Stays without labs went from
\MimAdReShareTwenty{} of the alerts to \MimAdMmShareTwenty, and xAUC fell from
\MimAdReXA{} to \MimAdMmXA, close to a coin flip. The catch rate fell in all \MimGrpN{}
race, age, sex and payer groups, by \MimGrpRelMin{} to \MimGrpRelMax{} of its value after
retraining. The largest relative loss was for \MimGrpWorst{} patients (\MimGrpWorstRe{}
to \MimGrpWorstMm), also the group least often tested by 3\,h here (62\% vs.\ 69\% for White patients), matching the Black--White testing gap Chang et al.\ found in
MIMIC-IV~\cite{chang2022disparate}. For ICU/death the loss
was smaller but clear. At top 10\% the catch rate without labs fell from
\MimIcuReSUTen{} to \MimIcuMmSUTen{} (\MimIcuMmVsReU\,pp [\MimIcuMmVsReULo,
\MimIcuMmVsReUHi]), and at top 5\% from \MimIcuReSUFive{} to \MimIcuMmSUFive.

Two thresholds fixed it here too (R4). On the admission list they caught
\MimAdTwoSUTwenty{} of admitted stays without labs (\CntMiTwo{} per test set; \MimAdTwoVsReU\,pp [\MimAdTwoVsReULo,
\MimAdTwoVsReUHi] against retraining) and \MimAdTwoSAllTwenty{} of all admitted stays,
a gain of \MimAdTwoVsReAll\,pp [\MimAdTwoVsReAllLo, \MimAdTwoVsReAllHi] over retraining,
and precision changed by \MimAdTwoVsRePPV\,pp. Within each of the \MimGrpN{} groups, the catch
rate without labs stayed within \MimGrpTwoMin{} to \MimGrpTwoMax\,pp of the retrain's.
For ICU/death, two thresholds brought the top-10\% catch rate to \MimIcuTwoSUTen,
slightly below retraining (\MimIcuTwoVsReU\,pp [\MimIcuTwoVsReULo, \MimIcuTwoVsReUHi]) but
well above the plain update, and the top-5\% rate to \MimIcuTwoSUFive. The additional checks also replicated. At a fixed risk cutoff the update flagged more stays but caught
\FmCutMmSU{} of admitted stays without labs against \FmCutReSU, and two cutoffs brought
it to \FmCutTwoSU. Even at twice the retrain's share, the overall catch rate
(\FmSwTwoAll) stayed above retraining. Among stays without labs, the update was again
better calibrated (ECE \FmEceMm{} against \FmEceRe; lower in all \FmEceGroups{} racial
groups).

\section{Discussion and Limitations}
At two hospitals, an update that adds useful data left patients without that data worse
off in two ways: with linear updates their order was reshuffled without new measurements, and
with every model they slipped down alert lists behind patients who were tested, even as the update caught more patients overall.
Whether that trade is acceptable is a clinical and ethical question to ask before deployment, all
the more because tests are ordered unevenly across groups. We suggest that teams report
flips and fixes among patients without the added data ($\CR_{UU}$, $\mathrm{IMP}_{UU}$)
and their alert-list catch rates, overall and by group, next to the usual aggregate
metrics.

These results add to earlier work. A compatibility loss makes updates more compatible at
little cost in AUROC~\cite{otles2023updating}, but compatibility over pairs did not protect
patients' place on an alert list. Retraining can affect groups unevenly~\cite{bilionis2026empirical};
adding a feature creates a new group, those without it, affected in a predictable direction.
Whether a test was done predicts outcomes~\cite{agniel2018biases,groenwold2020informative}, and
the update used that signal: better calibrated for untested patients, it still ranked patients
partly by whether they had been tested. An early alert may matter most for patients not yet
worked up, but that needs prospective study.

Two results surprised us. The alert loss came from the availability indicator in
Experiment~3 but from the lab values in Experiment~2, so dropping the indicator need not
suffice. And boosted trees mostly removed the extra reshuffling but not the alert loss.

The two-threshold policy needs someone to choose each group's share of alerts. We used
the retrain's share, which keeps things as they were, including any unfairness in how
alerts were shared before. That choice belongs to the clinical team, and the gap we still
saw for one group at MC-MED shows it should be checked within groups too. Next steps are
to repeat this analysis for deep fusion models and outside the US, to set the shares in a principled
way, for example with conformal risk control~\cite{angelopoulos2024crc} or
exposure-based rules~\cite{singh2018exposure}, and to ask clinicians how they react
when a patient's risk changes without any new measurement.

\subsection{Limitations}\label{sec:limits}
This is a retrospective study of simulated updates, so lower catch rates mean missed alerts, not
shown clinical harm. It covers two US academic centers, and at
both the added data that mattered was laboratory results. Our main updates are linear, and
Prop.~\ref{prop:uu} covers only that case. Boosted trees reproduced the alert loss but
little of the extra reshuffling, and we did not try deep multimodal fusion models. The null result of Experiment~1 holds only for our simple TF-IDF encoding of
the radiograph text. Our protocol is smaller than that
of~\cite{otles2023updating} (\NCand{} instead of 150 BCE candidates, 4 instead of 11
values of $\alpha$, \NReps{} instead of 40 repetitions), so we may not have explored the full trade-off between
compatibility and accuracy for RBC and MARBC. As there, the original model saw less data than the updates, which mainly affects the
retraining baseline.

\input{generated/tab_sens}
Refitting the lab block's imputation and scaling on training patients changed
almost nothing (Tab.~\ref{tab:sens}). Re-selecting the 25 components using only training patients changed 0--2
low-ranked ones per split; dropping them moved results by at most 0.1\,pp. Experiment~1's text vocabulary and SVD were not refit. Experiment~3's outcome, admission, is partly a clinical decision. At 3\,h,
clinicians may order laboratory tests \emph{because} they already expect to admit the
patient, so the update may partly learn their intentions from the ordering itself. That
is informative missingness too, but it means Experiment~3's alert effect is not only
about clinical deterioration. Experiment~2,
whose outcome is a clinical state, points the same way. At the second hospital we
skipped the ablations and the time-ordered split, and vital signs were charted rather than monitored.

Race is self-reported and includes a large Other/Unknown group. $\CR$ stands in for
clinicians' expectations~\cite{otles2023updating}, and we did not run a user study.
Finally, not every analysis was pre-specified. We stated Experiment~1's hypothesis
without a size threshold and designed Experiment~2 after seeing Experiment~1.
Experiment~3's hypotheses were recorded after the exploratory alert analysis of
Experiment~2 but before Experiment~3 was run, and the second hospital's before any
model was trained there. The two-threshold policy was exploratory at MC-MED; its MIMIC
admission evaluation under R4 was pre-specified. Analyses beyond the pre-specified
hypotheses are descriptive.

\section{Conclusion}
When a risk model gains a data source that only some patients have, the patients
without it do not stay put. In simulated updates at two hospitals, linear updates
re-ranked them more than retraining did when the new data helped prediction, and with
every model we tried they lost their place on alert lists to patients who were tested. A pre-specified
replication at a second hospital found a loss of almost the same size. Fixes that act on the model helped only partly; separate alert thresholds
for patients with and without the new data undid most of the loss at almost no cost.
Splitting compatibility by data availability, and reporting it with error correction
and alert-list effects, is a simple way to catch these changes before an update goes
live.

\section*{Data and Code Availability}
MC-MED is available to credentialed users through PhysioNet~\cite{kansal2025mcmeddata,
pollard2026physionet} under a data use agreement, and so are MIMIC-IV-ED and
MIMIC-IV~\cite{johnson2023mimicived,johnson2024mimiciv31}. Code and the recorded hypotheses are
available from the authors on request.

\bibliographystyle{IEEEtran}
\bibliography{refs}
\end{document}

%% file: generated/macros.tex
\newcommand{\NAllVisits}{118,385}
\newcommand{\NVisits}{78,475}
\newcommand{\NPatients}{50,323}
\newcommand{\OutcomeRate}{2.3\%}
\newcommand{\NPos}{1,822}
\newcommand{\NReps}{10}
\newcommand{\NCand}{9}
\newcommand{\Beta}{0.5}
\newcommand{\LamUU}{0.8}
\newcommand{\XRate}{22.6\%}
\newcommand{\XRateNeg}{22.0\%}
\newcommand{\XRatePos}{45.9\%}
\newcommand{\XChi}{272.4}
\newcommand{\XChiDof}{4}
\newcommand{\XChiP}{$p<0.001$}
\newcommand{\XMin}{19.2\%}
\newcommand{\XMinGroup}{Other/Unknown}
\newcommand{\XMax}{26.9\%}
\newcommand{\XMaxGroup}{NHPI}
\newcommand{\LRate}{83.4\%}

\newcommand{\LRatePos}{95.7\%}
\newcommand{\LChi}{163.9}
\newcommand{\LChiDof}{4}
\newcommand{\LChiP}{$p<0.001$}
\newcommand{\LMin}{79.4\%}
\newcommand{\LMinGroup}{Black}
\newcommand{\LMax}{84.9\%}
\newcommand{\LMaxGroup}{White}

\newcommand{\LHOneNSettings}{6}

\newcommand{\LHOneNSig}{4}

\newcommand{\TNReps}{6}
\newcommand{\TNPosU}{5}

\newcommand{\AFiveEHRBCESensU}{30.7\%}

\newcommand{\AFiveMMBCESensU}{22.7\%}

\newcommand{\AFiveDSensU}{$-$8.0}
\newcommand{\AFiveDSensULo}{$-$10.0}
\newcommand{\AFiveDSensUHi}{$-$5.9}

\newcommand{\AFiveDSensI}{+4.3}

\newcommand{\ANPosU}{24}

%% file: generated/macros_v3.tex
\newcommand{\VXAUROCe}{0.888}
\newcommand{\VXAUROCm}{0.888}
\newcommand{\VXFe}{3.5\%}
\newcommand{\VXFm}{3.5\%}
\newcommand{\VXX}{+0.04}
\newcommand{\VXXLo}{+0.02}
\newcommand{\VXXHi}{+0.05}

\newcommand{\VLFe}{6.0\%}
\newcommand{\VLFm}{7.2\%}
\newcommand{\VLX}{+1.18}
\newcommand{\VLXLo}{+0.60}
\newcommand{\VLXHi}{+1.76}

\newcommand{\VLRel}{20\%}

\newcommand{\VTX}{+5.05}
\newcommand{\VTXLo}{+0.75}
\newcommand{\VTXHi}{+9.36}

\newcommand{\VTDAu}{$-$0.0551}
\newcommand{\VTDAuLo}{$-$0.0924}
\newcommand{\VTDAuHi}{$-$0.0177}

\newcommand{\VGX}{+1.00}
\newcommand{\VGXLo}{$-$0.26}
\newcommand{\VGXHi}{+2.25}

\newcommand{\VAFe}{4.0\%}
\newcommand{\VAFm}{5.7\%}
\newcommand{\VAX}{+1.71}
\newcommand{\VAXLo}{+1.09}
\newcommand{\VAXHi}{+2.33}

\newcommand{\VADAu}{$-$0.0043}
\newcommand{\VADAuLo}{$-$0.0077}
\newcommand{\VADAuHi}{$-$0.0009}

\newcommand{\ETwentyReSensU}{18.8\%}

\newcommand{\ETwentyReRateU}{7.2\%}

\newcommand{\ETwentyMmSensU}{4.2\%}

\newcommand{\ETwentyMmRateU}{1.3\%}

\newcommand{\ETwentyRbcSensU}{11.3\%}

\newcommand{\ETwentyRtSensU}{13.9\%}

\newcommand{\ETwentyDSensU}{$-$14.6}
\newcommand{\ETwentyDSensULo}{$-$15.3}
\newcommand{\ETwentyDSensUHi}{$-$13.8}
\newcommand{\ETwentyDSensUP}{$p=0.002$}
\newcommand{\ETwentyDSensI}{+4.1}

\newcommand{\EAdmitRate}{44.8\%}
\newcommand{\EAdmitN}{35,127}
\newcommand{\ELabPos}{91.5\%}
\newcommand{\ELabNeg}{76.8\%}
\newcommand{\ETradeRatio}{3.0}

\newcommand{\EHOneNSettings}{6}

%% file: generated/macros_fu.tex
\newcommand{\DAFixUUe}{11.1\%}
\newcommand{\DAFixUUm}{14.2\%}

\newcommand{\DAFixIIe}{11.7\%}
\newcommand{\DAFixIIm}{34.5\%}

\newcommand{\HolmMax}{0.006}

\newcommand{\ETwoReSens}{30.7\%}

\newcommand{\EThrReAucU}{0.733}

\newcommand{\EThrReOE}{0.67}
\newcommand{\EThrRePred}{33.9\%}

\newcommand{\ETwoMmSens}{22.7\%}

\newcommand{\EThrMmAucU}{0.729}

\newcommand{\EThrMmOE}{0.85}
\newcommand{\EThrMmPred}{26.7\%}
\newcommand{\EThrMmObs}{22.7\%}

\newcommand{\ETwoNoFlagSens}{22.7\%}

\newcommand{\EThrNoFlagFlip}{5.7\%}
\newcommand{\EThrNoFlagSens}{11.6\%}

\newcommand{\EThrNoFlagDSens}{$-$7.2}
\newcommand{\EThrNoFlagDSensLo}{$-$8.0}
\newcommand{\EThrNoFlagDSensHi}{$-$6.5}

\newcommand{\ETwoFlagOnlySens}{29.3\%}

\newcommand{\EThrFlagOnlyAuc}{0.773}

\newcommand{\EThrFlagOnlyFlip}{4.1\%}
\newcommand{\EThrFlagOnlySens}{9.6\%}

\newcommand{\EThrFlagOnlyDSens}{$-$9.2}
\newcommand{\EThrFlagOnlyDSensLo}{$-$9.9}
\newcommand{\EThrFlagOnlyDSensHi}{$-$8.5}

\newcommand{\EThrNoiseAuc}{0.765}

\newcommand{\EThrNoiseFlip}{4.0\%}
\newcommand{\EThrNoiseSens}{18.7\%}

\newcommand{\EThrDropAuc}{0.797}

\newcommand{\EThrDropSens}{8.4\%}

\newcommand{\EThrAddAuc}{0.797}

\newcommand{\EThrAddSens}{7.3\%}

\newcommand{\GrpRelMin}{72\%}
\newcommand{\GrpRelMax}{80\%}
\newcommand{\GrpLowRe}{Black}
\newcommand{\GrpLowReV}{14.2\%}

\newcommand{\GrpN}{5}
\newcommand{\KReUTen}{8.0\%}

\newcommand{\KReUFifty}{55.4\%}
\newcommand{\KMmUTen}{0.8\%}

\newcommand{\KMmUFifty}{35.9\%}

%% file: generated/macros_f2.tex
\newcommand{\PTNReps}{6}

\newcommand{\PAReSU}{18.8\%}

\newcommand{\PAReSAll}{35.9\%}

\newcommand{\PARePPV}{80.2\%}
\newcommand{\PAReXA}{0.636}

\newcommand{\PAReFlip}{4.0\%}
\newcommand{\PTReSU}{16.4\%}

\newcommand{\PLReSU}{30.7\%}

\newcommand{\PAMmSAll}{38.4\%}

\newcommand{\PAMmPPV}{85.8\%}
\newcommand{\PAMmXA}{0.532}

\newcommand{\PTMmSU}{3.5\%}

\newcommand{\PTMmSAll}{39.4\%}

\newcommand{\PLMmSU}{22.7\%}

\newcommand{\PLMmSAll}{52.5\%}

\newcommand{\PATwoSU}{18.7\%}

\newcommand{\PATwoSAll}{38.1\%}

\newcommand{\PATwoPPV}{85.0\%}
\newcommand{\PTTwoSU}{15.2\%}

\newcommand{\PTTwoSAll}{39.0\%}

\newcommand{\PLTwoSU}{27.8\%}

\newcommand{\PLTwoSAll}{52.4\%}

\newcommand{\PAAncSU}{6.4\%}

\newcommand{\PAAncAuc}{0.799}
\newcommand{\PAAncFlip}{4.1\%}

\newcommand{\PATwoVsReU}{$-$0.1}
\newcommand{\PATwoVsReULo}{$-$1.5}
\newcommand{\PATwoVsReUHi}{+1.2}

\newcommand{\PTMmVsReU}{$-$12.9}
\newcommand{\PTMmVsReULo}{$-$13.2}
\newcommand{\PTMmVsReUHi}{$-$12.6}
\newcommand{\PTMmVsReFlip}{+2.4}
\newcommand{\PTMmVsReFlipLo}{+2.2}
\newcommand{\PTMmVsReFlipHi}{+2.5}

\newcommand{\CoefAgeRe}{0.55}
\newcommand{\CoefAgeMm}{0.41}
\newcommand{\CoefShrink}{55\%}

\newcommand{\SgRelMin}{74\%}
\newcommand{\SgRelMax}{88\%}
\newcommand{\SgWorst}{Medicaid}
\newcommand{\SgWorstRe}{6.3\%}
\newcommand{\SgWorstMm}{0.7\%}
\newcommand{\SgN}{8}
\newcommand{\LabAgeYoung}{79\%}
\newcommand{\LabAgeOld}{88\%}
\newcommand{\TwoGrpN}{13}
\newcommand{\TwoGrpWorst}{Black}
\newcommand{\TwoGrpWorstD}{$-$2.5}
\newcommand{\TwoGrpWorstLo}{$-$4.5}
\newcommand{\TwoGrpWorstHi}{$-$0.6}
\newcommand{\TwoGrpWorstRe}{14.2\%}
\newcommand{\TwoGrpWorstTwo}{11.7\%}
\newcommand{\TwoGrpWorstMm}{3.9\%}
\newcommand{\TwoGrpNBelow}{one}
\newcommand{\TwoGrpOthMin}{$-$1.2}
\newcommand{\TwoGrpOthMax}{+3.5}

%% file: generated/macros_mimic.tex
\newcommand{\MimAll}{347{,}036}
\newcommand{\MimPatients}{171{,}701}
\newcommand{\MimN}{100{,}000}
\newcommand{\MimIcuRate}{6.2\%}
\newcommand{\MimIcuLabPos}{89\%}

\newcommand{\MimAdRate}{41.0\%}
\newcommand{\MimAdLabPos}{81\%}
\newcommand{\MimAdLabNeg}{57\%}
\newcommand{\MimLab}{67\%}
\newcommand{\MimLabMin}{62\%}
\newcommand{\MimLabMinGroup}{Black}
\newcommand{\MimLabMax}{69\%}
\newcommand{\MimLabMaxGroup}{White}

\newcommand{\MimIcuFlipD}{+0.57}
\newcommand{\MimIcuFlipDLo}{+0.37}
\newcommand{\MimIcuFlipDHi}{+0.77}

\newcommand{\MimAdFlipD}{+1.22}
\newcommand{\MimAdFlipDLo}{+0.83}
\newcommand{\MimAdFlipDHi}{+1.61}

\newcommand{\MimAdDAucU}{$-$0.003}
\newcommand{\MimAdDAucULo}{$-$0.004}
\newcommand{\MimAdDAucUHi}{$-$0.002}
\newcommand{\MimAdFixUURe}{12\%}
\newcommand{\MimAdFixUUMm}{14\%}
\newcommand{\MimAdFixIIRe}{11\%}
\newcommand{\MimAdFixIIMm}{36\%}

\newcommand{\MimIcuReSUFive}{23.4\%}

\newcommand{\MimIcuReSUTen}{38.4\%}

\newcommand{\MimIcuMmSUFive}{14.8\%}

\newcommand{\MimIcuMmSUTen}{31.2\%}

\newcommand{\MimIcuTwoSUFive}{23.8\%}

\newcommand{\MimIcuTwoSUTen}{37.1\%}

\newcommand{\MimIcuMmVsReU}{$-$7.3}
\newcommand{\MimIcuMmVsReULo}{$-$8.9}
\newcommand{\MimIcuMmVsReUHi}{$-$5.7}
\newcommand{\MimIcuTwoVsReU}{$-$1.3}
\newcommand{\MimIcuTwoVsReULo}{$-$2.4}
\newcommand{\MimIcuTwoVsReUHi}{$-$0.2}

\newcommand{\MimAdReSUTwenty}{22.6\%}

\newcommand{\MimAdReShareTwenty}{13.5\%}

\newcommand{\MimAdReXA}{0.65}

\newcommand{\MimAdMmSUTwenty}{4.9\%}

\newcommand{\MimAdMmShareTwenty}{2.3\%}

\newcommand{\MimAdMmXA}{0.52}

\newcommand{\MimAdTwoSUTwenty}{22.4\%}
\newcommand{\MimAdTwoSAllTwenty}{40.1\%}

\newcommand{\MimAdMmVsReU}{$-$17.7}
\newcommand{\MimAdMmVsReULo}{$-$18.3}
\newcommand{\MimAdMmVsReUHi}{$-$17.1}
\newcommand{\MimAdTwoVsReU}{$-$0.2}
\newcommand{\MimAdTwoVsReULo}{$-$1.1}
\newcommand{\MimAdTwoVsReUHi}{+0.8}
\newcommand{\MimAdTwoVsReAll}{+2.2}
\newcommand{\MimAdTwoVsReAllLo}{+2.2}
\newcommand{\MimAdTwoVsReAllHi}{+2.3}

\newcommand{\MimAdTwoVsRePPV}{+4.6}

\newcommand{\MimHolmMax}{0.008}
\newcommand{\MimGrpN}{14}
\newcommand{\MimGrpRelMin}{70\%}
\newcommand{\MimGrpRelMax}{89\%}
\newcommand{\MimGrpWorst}{Black}
\newcommand{\MimGrpWorstRe}{14.9\%}
\newcommand{\MimGrpWorstMm}{1.6\%}
\newcommand{\MimGrpTwoMin}{$-$1.9}
\newcommand{\MimGrpTwoMax}{+1.4}
\newcommand{\MimCoefShrink}{67 of 76}
\newcommand{\MimCoefAgeRe}{0.60}
\newcommand{\MimCoefAgeMm}{0.44}
\newcommand{\CntMcN}{23{,}600}
\newcommand{\CntMcPos}{889}
\newcommand{\CntMcRe}{167}
\newcommand{\CntMcMm}{38}

\newcommand{\CntMiN}{30{,}000}
\newcommand{\CntMiPos}{2{,}260}
\newcommand{\CntMiRe}{511}
\newcommand{\CntMiMm}{111}
\newcommand{\CntMiTwo}{506}

%% file: generated/macros_f3.tex
\newcommand{\FcSwZeroAll}{38.4\%}
\newcommand{\FcSwTwoAll}{37.2\%}
\newcommand{\FcSwTwoU}{32.2\%}

\newcommand{\FcSwMinPPV}{83.1\%}
\newcommand{\FcCutRePct}{20.0\%}
\newcommand{\FcCutReSU}{18.8\%}

\newcommand{\FcCutMmPct}{23.2\%}
\newcommand{\FcCutMmSU}{5.8\%}

\newcommand{\FcCutTwoSU}{18.7\%}

\newcommand{\FcEceRe}{0.112}
\newcommand{\FcEceMm}{0.040}

\newcommand{\FcEceGroups}{five}

\newcommand{\FmSwTwoAll}{38.3\%}

\newcommand{\FmCutReSU}{22.3\%}

\newcommand{\FmCutMmSU}{6.7\%}

\newcommand{\FmCutTwoSU}{22.2\%}

\newcommand{\FmEceRe}{0.091}
\newcommand{\FmEceMm}{0.036}

\newcommand{\FmEceGroups}{five}

%% file: generated/macros_gbm2.tex
\newcommand{\GcFlip}{$-$0.04}
\newcommand{\GcFlipLo}{$-$0.34}
\newcommand{\GcFlipHi}{+0.27}

\newcommand{\GcReSU}{20.3\%}

\newcommand{\GcMmSU}{6.0\%}

\newcommand{\GcTwoSU}{20.3\%}

\newcommand{\GcDSU}{$-$14.3}
\newcommand{\GcDSULo}{$-$15.3}
\newcommand{\GcDSUHi}{$-$13.3}
\newcommand{\GmFlip}{$-$0.40}
\newcommand{\GmFlipLo}{$-$0.63}
\newcommand{\GmFlipHi}{$-$0.18}

\newcommand{\GmReSU}{24.3\%}

\newcommand{\GmMmSU}{9.5\%}

\newcommand{\GmTwoSU}{24.4\%}

\newcommand{\GiFlip}{+0.39}
\newcommand{\GiFlipLo}{+0.14}
\newcommand{\GiFlipHi}{+0.64}

%% file: figs/fig_concept.tex
\begin{tikzpicture}[font=\scriptsize, x=1cm, y=1cm,
  txt/.style={align=center, inner sep=0pt},
  hdr/.style={align=center, text=black!70}]
  \def\W{2.6}\def\H{1.1}
  \fill[black!4]  (\W,0) rectangle (2*\W,\H);
  \fill[black!4]  (0,-\H) rectangle (\W,0);
  \fill[black!12] (\W,-\H) rectangle (2*\W,0);
  \fill[orange!15] (0,0) rectangle (\W,\H);
  \draw[black!50, line width=0.4pt] (0,-\H) grid[xstep=\W,ystep=\H] (2*\W,\H);
  \draw[orange!75!black, line width=0.9pt] (0,0) rectangle (\W,\H);
  \node[txt] at (\W/2,\H/2)      {\textbf{UU}\\[1pt] neither has labs\\ \textit{no new evidence}};
  \node[txt] at (1.5*\W,\H/2)    {\textbf{Mixed}\\[1pt] only this patient\\ has labs};
  \node[txt] at (\W/2,-\H/2)     {\textbf{Mixed}\\[1pt] only this patient\\ has labs};
  \node[txt] at (1.5*\W,-\H/2)   {\textbf{II}\\[1pt] both have labs\\ \textit{new evidence}};
  \node[hdr] at (\W/2,\H+0.2)    {no labs};
  \node[hdr] at (1.5*\W,\H+0.2)  {labs};
  \node[font=\scriptsize\bfseries] at (\W,\H+0.55) {Patient with the outcome};
  \node[hdr, rotate=90] at (-0.2,\H/2)  {no labs};
  \node[hdr, rotate=90] at (-0.2,-\H/2) {labs};
  \node[font=\scriptsize\bfseries, rotate=90] at (-0.55,0) {Patient without it};
\end{tikzpicture}

%% file: generated/tab_cohort.tex
\begin{table}[t]\centering\small
\caption{Cohort and availability of each added source by 3\,h, by race (neg/pos: without/with the outcome).}
\label{tab:cohort}
\begin{adjustbox}{max width=\linewidth}\begin{tabular}{lrrrrrrr}\toprule
 & & & \multicolumn{2}{c}{CXR} & & \multicolumn{2}{c}{Labs}\\
\cmidrule(lr){4-5}\cmidrule(lr){7-8}
Group & $n$ & Outcome & all & neg/pos & & all & neg/pos\\\midrule
Asian & 12,841 & 0.025 & 0.238 & 0.23/0.47 & & 0.844 & 0.84/0.95\\
Black & 5,004 & 0.025 & 0.233 & 0.23/0.46 & & 0.794 & 0.79/0.97\\
NHPI & 1,537 & 0.027 & 0.269 & 0.26/0.60 & & 0.833 & 0.83/1.00\\
Other/Unknown & 26,974 & 0.018 & 0.192 & 0.19/0.44 & & 0.819 & 0.82/0.95\\
White & 32,119 & 0.026 & 0.245 & 0.24/0.46 & & 0.849 & 0.85/0.96\\
\midrule
All & 78,475 & 0.023 & 0.226 & 0.22/0.46 & & 0.834 & 0.83/0.96\\
\bottomrule\end{tabular}\end{adjustbox}\end{table}

%% file: generated/tab_summary.tex
\begin{table*}[t]\centering\footnotesize\setlength{\tabcolsep}{4pt}
\caption{Patients without the added data: retrain $\to$ update; differences mean [95\% CI]. $^\ast$Pre-specified.}
\label{tab:summary}
\begin{tabular}{clllcccc}\toprule
Exp. & Added data & Outcome, design & AUROC & \multicolumn{2}{c}{Flipped (\%)} & Extra flips (pp) & $\Delta$AUROC$_{\varnothing}$\\
\cmidrule(lr){5-6}
 & & & retrain$\to$update & retrain & update & & \\\midrule
1 & Radiograph text & ICU/death, random & 0.888$\to$0.888 & 3.5 & 3.5 & \textbf{+0.04 [+0.02, +0.05]} & $-$0.0001 [$-$0.0003, 0.0000]\\
2 & Lab results & ICU/death, random$^\ast$ & 0.868$\to$0.885 & 6.0 & 7.2 & \textbf{+1.18 [+0.60, +1.76]} & $-$0.0042 [$-$0.0106, +0.0023]\\
2 & Lab results & ICU/death, time-ordered & 0.852$\to$0.882 & 1.3 & 6.4 & \textbf{+5.05 [+0.75, +9.36]} & \textbf{$-$0.0551 [$-$0.0924, $-$0.0177]}\\
2 & Lab results & ICU/death, boosting & 0.870$\to$0.891 & 8.7 & 9.7 & +1.00 [$-$0.26, +2.25] & $-$0.0095 [$-$0.0228, +0.0037]\\
3 & Lab results & admission, random$^\ast$ & 0.767$\to$0.802 & 4.0 & 5.7 & \textbf{+1.71 [+1.09, +2.33]} & \textbf{$-$0.0043 [$-$0.0077, $-$0.0009]}\\
\bottomrule\end{tabular}\end{table*}

%% file: generated/tab_decomp.tex
\begin{table}[t]\centering
\caption{Flipped and fixed pairs by pair type (\%), retrain $\to$ update.}
\label{tab:decomp}
\begin{adjustbox}{max width=\columnwidth}
\begin{tabular}{llrrr}\toprule
Exp. & & UU & II & Mixed\\\midrule
2 & flipped & 6.0$\to$7.2 & 3.0$\to$3.6 & 2.7$\to$3.0\\
  & fixed & 23.3$\to$25.6 & 24.4$\to$41.2 & 24.7$\to$36.3\\
3 & flipped & 4.0$\to$5.7 & 3.4$\to$5.8 & 3.1$\to$4.9\\
  & fixed & 11.1$\to$14.2 & 11.7$\to$34.5 & 11.9$\to$34.0\\
\bottomrule\end{tabular}
\end{adjustbox}
\end{table}

%% file: generated/tab_ablation.tex
\begin{table*}[t]\centering
\caption{Variants of the update (Caught: top 5\% in Exp.~2, top 20\% in Exp.~3).}
\label{tab:ablation}
\begin{adjustbox}{max width=\textwidth}
\begin{tabular}{lrrrrrrrr}\toprule
 & \multicolumn{4}{c}{Exp.~2: ICU/death, top 5\%} & \multicolumn{4}{c}{Exp.~3: admission, top 20\%}\\
\cmidrule(lr){2-5}\cmidrule(lr){6-9}
Model & AUROC & Flipped & Caught & O/E & AUROC & Flipped & Caught & O/E\\\midrule
EHR-only retrain & 0.868 & 6.0 & 30.7 & 0.58 & 0.767 & 4.0 & 18.8 & 0.67\\
Update: labs + indicator & 0.885 & 7.2 & 22.7 & 0.72 & 0.802 & 5.7 & 4.2 & 0.85\\
\addlinespace[2pt]
\quad labs, no indicator & 0.885 & 7.2 & 22.7 & 0.70 & 0.793 & 5.7 & 11.6 & 0.64\\
\quad indicator, no labs & 0.869 & 6.3 & 29.3 & 0.61 & 0.773 & 4.1 & 9.6 & 0.86\\
\quad random features, no indicator & 0.864 & 6.3 & 30.2 & 0.60 & 0.765 & 4.0 & 18.7 & 0.67\\
\quad modality dropout (30\%) & 0.882 & 6.5 & 24.2 & 0.63 & 0.797 & 4.6 & 8.4 & 0.71\\
\quad frozen original + additive head & 0.876 & 0.0 & 20.4 & 0.77 & 0.797 & 0.0 & 7.3 & 0.86\\
\quad RBC, probability scale & 0.884 & 5.7 & 22.5 & 0.37 & 0.785 & 1.2 & 13.5 & 0.75\\
\bottomrule\end{tabular}
\end{adjustbox}
\end{table*}

%% file: generated/tab_groups.tex
\begin{table}[t]\centering
\caption{Experiment~3, top-20\% list: admitted visits without labs caught, by race (n per test set).}
\label{tab:groups}
\begin{adjustbox}{max width=\columnwidth}
\begin{tabular}{lrrrlr}\toprule
 & & \multicolumn{3}{c}{Without labs} & With labs\\\cmidrule(lr){3-5}\cmidrule(lr){6-6}
Group & $n$ & Retrain & Update & Difference [95\% CI] & Difference\\\midrule
Asian & 139 & 21.2 & 4.3 & $-$16.9 [$-$18.4, $-$15.4] & +4.6\\
Black & 63 & 14.2 & 3.9 & $-$10.3 [$-$13.7, $-$6.9] & +6.4\\
NHPI & 15 & 19.0 & 5.4 & $-$13.6 [$-$19.0, $-$8.2] & +6.5\\
Other/Unknown & 300 & 17.5 & 3.7 & $-$13.8 [$-$15.3, $-$12.4] & +6.1\\
White & 373 & 19.6 & 4.7 & $-$14.9 [$-$15.7, $-$14.1] & +2.2\\
\bottomrule\end{tabular}
\end{adjustbox}
\end{table}

%% file: generated/tab_mitig.tex
\begin{table*}[t]\centering\footnotesize\setlength{\tabcolsep}{4pt}

\caption{Mitigations minus the plain update, mean [95\% CI]; bold: CI excludes 0. All in pp except AUROC.}
\label{tab:mitig}
\begin{tabular}{llcccc}\toprule
Exp. & Mitigation & $\Delta$AUROC & $\Delta$Flipped (pp) & $\Delta$Fixed (pp) & $\Delta$Caught (pp)\\\midrule
1 (radiograph text) & RBC~\cite{otles2023updating} & $-$0.0005 [$-$0.0025, +0.0015] & \textbf{$-$1.96 [$-$2.31, $-$1.62]} & \textbf{$-$11.6 [$-$13.7, $-$9.6]} & --\\
 & Routing & \textbf{$-$0.0093 [$-$0.0123, $-$0.0064]} & \textbf{$-$3.55 [$-$3.97, $-$3.13]} & \textbf{$-$32.7 [$-$36.2, $-$29.2]} & --\\
\midrule
2 (labs, ICU/death) & RBC~\cite{otles2023updating} & \textbf{$-$0.0033 [$-$0.0060, $-$0.0006]} & \textbf{$-$3.58 [$-$4.65, $-$2.51]} & \textbf{$-$6.7 [$-$9.7, $-$3.6]} & $-$0.2 [$-$2.8, +2.4]\\
 & Routing & $-$0.0009 [$-$0.0020, +0.0002] & \textbf{$-$7.17 [$-$8.50, $-$5.84]} & \textbf{$-$25.6 [$-$31.9, $-$19.4]} & +2.4 [$-$1.3, +6.2]\\
\midrule
3 (labs, admission) & RBC~\cite{otles2023updating} & \textbf{$-$0.0124 [$-$0.0131, $-$0.0117]} & \textbf{$-$3.79 [$-$4.23, $-$3.35]} & \textbf{$-$8.6 [$-$9.5, $-$7.7]} & \textbf{+7.1 [+6.6, +7.6]}\\
 & Routing & \textbf{$-$0.0040 [$-$0.0046, $-$0.0034]} & \textbf{$-$5.72 [$-$6.27, $-$5.18]} & \textbf{$-$14.2 [$-$15.4, $-$13.1]} & \textbf{+9.7 [+8.9, +10.5]}\\
\bottomrule\end{tabular}\end{table*}

%% file: generated/tab_policy.tex
\begin{table*}[t]\centering
\caption{Alert lists under the two-threshold policy.}
\label{tab:policy}
\begin{adjustbox}{max width=\textwidth}
\begin{tabular}{lrrrrrrrrrrrrrrrrrr}\toprule
 & \multicolumn{6}{c}{Exp.~3, admission, top 20\%} & \multicolumn{6}{c}{Exp.~3, time-ordered, top 20\%} & \multicolumn{6}{c}{Exp.~2, ICU/death, top 5\%}\\
\cmidrule(lr){2-7}\cmidrule(lr){8-13}\cmidrule(lr){14-19}
Model & U & I & All & Prec. & Share U & xAUC & U & I & All & Prec. & Share U & xAUC & U & I & All & Prec. & Share U & xAUC\\\midrule
EHR-only retrain & 18.8 & 37.5 & 35.9 & 80.2 & 5.9 & 0.636 & 16.4 & 38.3 & 36.5 & 77.7 & 5.0 & 0.646 & 30.7 & 49.6 & 48.7 & 22.7 & 3.0 & 0.669\\
Multimodal update & 4.2 & 41.6 & 38.4 & 85.8 & 1.0 & 0.532 & 3.5 & 42.6 & 39.4 & 83.9 & 1.0 & 0.538 & 22.7 & 53.9 & 52.5 & 24.5 & 1.7 & 0.660\\
\quad + two thresholds & 18.7 & 39.8 & 38.1 & 85.0 & 6.0 & -- & 15.2 & 41.2 & 39.0 & 83.1 & 4.7 & -- & 27.8 & 53.6 & 52.4 & 24.4 & 2.8 & --\\
\quad anchored to retrain & 6.4 & 41.1 & 38.2 & 85.3 & 1.6 & 0.553 & 5.4 & 42.0 & 38.9 & 83.0 & 1.4 & 0.564 & 24.2 & 53.8 & 52.5 & 24.5 & 1.7 & 0.666\\
\bottomrule\end{tabular}
\end{adjustbox}
\end{table*}

%% file: generated/tab_mimic.tex
\begin{table}[t]\centering
\caption{Second hospital (MIMIC-IV-ED, pre-specified).}
\label{tab:mimic}
\begin{adjustbox}{max width=\columnwidth}
\begin{tabular}{lrrrrrrr}\toprule
Model & AUROC & AUROC$_\varnothing$ & Flipped & U & All & Prec. & xAUC\\\midrule
\multicolumn{8}{l}{\emph{ICU/death, top 10\%}}\\
EHR-only retrain & 0.888 & 0.850 & 2.3 & 38.4 & 62.1 & 38.3 & 0.74\\
Multimodal update & 0.903 & 0.847 & 2.9 & 31.2 & 65.2 & 40.2 & 0.72\\
\quad + two thresholds & -- & -- & -- & 37.1 & 65.0 & 40.1 & --\\
\midrule
\multicolumn{8}{l}{\emph{Admission, top 20\%}}\\
EHR-only retrain & 0.768 & 0.758 & 3.4 & 22.6 & 37.8 & 77.2 & 0.65\\
Multimodal update & 0.805 & 0.755 & 4.6 & 4.9 & 40.6 & 82.8 & 0.52\\
\quad + two thresholds & -- & -- & -- & 22.4 & 40.1 & 81.8 & --\\
\bottomrule\end{tabular}
\end{adjustbox}
\end{table}

%% file: generated/tab_sens.tex
\begin{table}[t]\centering
\caption{Lab preprocessing on all visits (Orig.) vs.\ training patients only (Train).}
\label{tab:sens}
\begin{adjustbox}{max width=\columnwidth}
\begin{tabular}{lrrrrrr}\toprule
& \multicolumn{2}{c}{Extra flips (pp)} & \multicolumn{2}{c}{Update (\%)} & \multicolumn{2}{c}{2 thr.\ (\%)}\\
\cmidrule(lr){2-3}\cmidrule(lr){4-5}\cmidrule(lr){6-7}
Setting & Orig. & Train & Orig. & Train & Orig. & Train\\\midrule
Exp.~2, ICU/death, top 5\% & +1.18 & +1.17 & 22.7 & 22.7 & 27.8 & 27.8\\
Exp.~3, admission, top 20\% & +1.71 & +1.71 & 4.2 & 4.2 & 18.7 & 18.7\\
MIMIC, ICU/death, top 10\% & +0.57 & +0.57 & 31.2 & 31.1 & 37.1 & 37.1\\
MIMIC, admission, top 20\% & +1.22 & +1.22 & 4.9 & 4.9 & 22.4 & 22.4\\
\bottomrule\end{tabular}
\end{adjustbox}
\end{table}